\documentclass[10pt,letterpaper]{article}
\usepackage{opex3}
\usepackage{amssymb,amsmath}
\usepackage{cite}

\begin{document}

\title{A chip-scale, telecommunications-band frequency conversion interface for quantum emitters\hspace{5mm} }

\author{Imad Agha,$^{1,2,4,5}$ Serkan Ates,$^{1,2,4}$ Marcelo Davan\c{c}o,$^{1,3}$ and Kartik Srinivasan$^{1,\ast}$}

\address{$^1$Center for Nanoscale Science and Technology, National
Institute of Standards and Technology, Gaithersburg, MD 20899\\
$^2$Maryland NanoCenter, University of Maryland, College Park, MD
20742\\
$^3$Department of Applied Physics, California Institute of Technology, Pasadena, CA 91125\\
$^{4}$These authors contributed equally.\\
$^{5}${imad.agha@nist.gov}\\ 
$^{\ast}${kartik.srinivasan@nist.gov}}

\begin{abstract} We describe a chip-scale, telecommunications-band frequency conversion interface
designed for low-noise operation at wavelengths desirable for common single photon emitters.
Four-wave-mixing Bragg scattering in silicon nitride waveguides is used to demonstrate frequency upconversion and
downconversion between the 980~nm and 1550~nm wavelength regions, with
signal-to-background levels $>10$ and conversion efficiency of $\approx -60$ dB
at low continuous wave input pump powers ($<50$~mW).  Finite element simulations and the
split-step Fourier method indicate that increased input powers of $\approx$10~W
(produced by amplified nanosecond pulses, for example) will result in a conversion
efficiency $>25~\%$ in existing geometries.  Finally, we present waveguide designs that can
be used to connect shorter wavelength (637~nm to 852~nm) quantum emitters with 1550~nm. \end{abstract}




\section{Introduction}

Interfaces that connect photonic quantum systems operating at different
frequencies constitute an important resource in the development of a hybrid
architecture that leverages the key advantages of its component entities.
Such hybrid architectures may be at the heart of modern quantum networks,
which must be able to generate, send, manipulate, and store quantum
information with high fidelity and low loss.  For example, single photon
sources (SPSs) based on single quantum emitters like InAs quantum
dots~\cite{ref:Michler_book_2009}, nitrogen vacancy centers in
diamond~\cite{ref:Kurtsiefer}, and neutral alkali atoms~\cite{ref:McKeever}
all exhibit desirable features, such as on-demand generation with the
potential for high single photon purity and indistinguishability. These
properties are crucial in both quantum cryptography~\cite{ref:Grangier_QKD}
and quantum computing~\cite{ref:Knill,ref:OBrien_Furusawa_Vuckovic}
applications. Unfortunately, for these common quantum emitters, emission
occurs at wavelengths below 1000~nm, where long-distance transmission through
optical fibers is not optimal. Similarly, many promising quantum
memories~\cite{ref:Simon_Qmemory_review} operate in the visible or
near-visible, whereas silicon-based systems that have been developed for
applications in nanophotonics, CMOS electronics, and microelectromechanics
are opaque at such wavelengths.  Thus, interfaces that can connect disparate
wavelength regions without otherwise disturbing the relevant properties of the quantum state
(such as photon statistics and coherence time) are an important
resource for future photonic quantum information processing systems.  Such quantum frequency
conversion~\cite{ref:Kumar_OL,ref:Raymer_Srinivasan_PT} has been demonstrated
with single photon Fock states generated by single semiconductor quantum
dots~\cite{ref:Rakher_NPhot_2010,ref:Zaske_Becher_downconversion,ref:Ates_Srinivasan_PRL,ref:de_Greve_Yamamoto_Nature},
where three-wave-mixing in periodically-poled lithium niobate waveguides was
used for both
upconversion~\cite{ref:Rakher_NPhot_2010,ref:Ates_Srinivasan_PRL} and
downconversion~\cite{ref:Zaske_Becher_downconversion,ref:de_Greve_Yamamoto_Nature}.

Future applications may benefit from the development of such frequency
conversion technology within silicon-based material systems, for which
complex and highly scalable fabrication technology at a level beyond lithium
niobate has been demonstrated.  To that end, frequency conversion through
four-wave-mixing Bragg scattering (FWM-BS)~\cite{ref:McKinstrie}, a process
used in quantum frequency conversion experiments in optical
fibers~\cite{ref:McGuinnes_PRL10}, was recently demonstrated in silicon
nitride (Si$_3$N$_4$) waveguides~\cite{ref:Agha_OL}, where Si$_3$N$_4$ was chosen due to its
wide optical transparency window (including visible and near-visible wavelengths of importance to
many quantum optical systems), its relatively high nonlinear refractive index, and low linear and
nonlinear absorption. In Ref.~\cite{ref:Agha_OL}, narrowband conversion over a few nanometers was measured, and the
low-noise nature was inferred by measuring the pump-induced noise.  Here, we show the
flexibility of FWM-BS in these Si$_3$N$_4$ devices by demonstrating wideband downconversion and upconversion between the 980~nm and 1550~nm
bands, with a signal-to-noise ratio $\gtrsim10$ as determined via photon
counting measurements.  Through calculations of waveguide dispersion and simulations using
the split-step Fourier method, we further show that simple variations in the waveguide geometry should extend
this technique to enable quantum emitters in the 637~nm, 780~nm, and 850~nm
wavelength bands to be connected to the 1550~nm band. Finally, as the demonstrated conversion
efficiencies are low ($\approx-60$~dB) due to relatively low pump power used ($<50$~mW), we use
simulations to establish the pump powers required to achieve conversion efficiencies high enough for
realistic experiments with single photon states of light.  We find that pump powers of $\approx$10~W, achievable
through amplified nanosecond pulses, for example, should enable conversion efficiencies $>25~\%$ to be achieved.

\section{Principle of operation}

\begin{figure}[h]
\begin{center}
\centerline{\includegraphics[width=\linewidth,trim= 0 0 0
0]{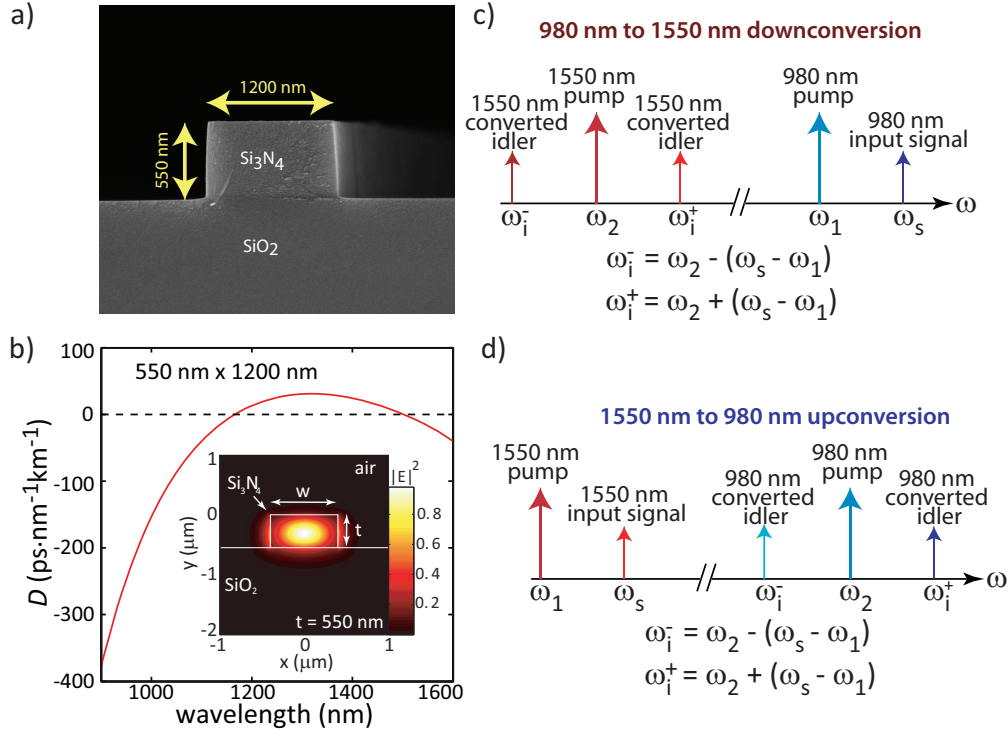}} \caption{a) Scanning electron microscope image of the 550~nm x 1200~nm
silicon nitride waveguide used in the current experiment. b) Dispersion parameter $D=\frac{-2\pi c}{\lambda^2}\frac{d^2\beta}{d\omega^2}$
as a function of wavelength $\lambda$ ($\beta$ is the propagation constant, $c$ is the speed of light). The zero-dispersion point is around 1200~nm,
as required by phase-matching considerations. The inset shows the numerically calculated electric field amplitude for the fundamental TE waveguide mode. c) and d)  Position of the pumps, signal, and idlers in frequency space for frequency downconversion (c) and upconversion (d).
} \label{fig:waveguide and schematics}
\end{center}
\end{figure}

Four-wave-mixing Bragg scattering is a nonlinear mixing process involving
four non-degenerate fields, whereby two pumps $\omega_1$  and $\omega_2$
mediate the scattering from a signal $\omega_s$ to an idler $\omega_i$.
 In this paper, we consider a process in which the idler frequencies are given by
 $\omega_{i}^{\pm}=\omega_2 \pm (\omega_s - \omega_1)$. In the slowly-varying envelope
approximation, and  ignoring the effects of  pump-depletion, the conversion
efficiency from the signal to the idler  at a propagation distance $z$ along
the waveguide is given by:

\begin{equation}
\eta (z) = \frac{4\gamma_1 \gamma_2 P_1 P_2}{g^2} \sin^2(gz)
\end{equation}
where $\gamma_{1}$ and $\gamma_{2}$ are the waveguide's effective nonlinearity parameter at the two
pump wavelengths, $P_1$ and $P_2$ are the powers in pumps 1 and 2,
respectively, and $g$ is the four-wave-mixing gain that depends both on the
pump powers (through nonlinear dispersion) and the linear phase
mismatch~\cite{ref:Uesaka_Kazovksy}. A typical estimated value for the effective nonlinearity parameter is
$\gamma\approx6$~W$^{-1}$m$^{-1}$ (depending on the specific waveguide cross-section and wavelength), which
is determined assuming a nonlinear refractive index $n_{2}=2.5\times10^{-19}$~m$^2$W$^{-1}$\cite{ref:Ikeda_Fainman_silicon_nitride_nonlinear_OE}. By tailoring the dispersion
appropriately, and by having the two pumps widely separated in wavelength,
this process allows ultra-wide band frequency conversion. Moreover, due to
the fact that this is a wavelength exchange process with no net amplification
of either signal or idler, the conversion is theoretically noiseless and is
suitable for quantum state translation~\cite{ref:McKinstrie}. In reality, the
noise limit may come either from poor filtering of the pumps or, in certain
materials, from Raman scattering or other four-wave-mixing processes.

In this work, our main motivation is to develop an interface for frequency
conversion from 980~nm to 1550~nm (InAs quantum dot SPSs to telecom
wavelength) in a silicon nitride waveguide (Fig.~\ref{fig:waveguide and
schematics}(a)). Silicon nitride is chosen here due to CMOS fabrication
compatibility~\cite{ref:Levy_Lipson_comb}, simple growth techniques, large
transparency window, high optical nonlinearity (x10 that of
SiO$_2$)~\cite{ref:Tan_Fainman}, and low nonlinear loss (absence of
two-photon absorption and free-carrier generation). Details on the
fabrication technique and dispersion calculations (Fig.~\ref{fig:waveguide
and schematics}(b)) can be found in Ref.~\cite{ref:Agha_OL}.    Unlike this
earlier  work, whereby we employed two telecom-band pumps to convert a signal
to its idler in the 980~nm band, here we place one of pumps near 1550~nm and
the other around 980~nm (Fig.~\ref{fig:waveguide and schematics}(c,d)),
allowing for downconversion (upconversion) over a $\approx 600$~nm range, a
significantly larger wavelength translation range compared to wideband
conversion recently demonstrated in optical fibers through
FWM-BS~\cite{ref:McGuinness_PTL}.  As the nonlinear phase-mismatch scales as
$\Delta\beta_{nl} = \gamma_1P_1-\gamma_2P_2$, where $\gamma_{1}$ ($\gamma_{2}$) is the
nonlinear coefficient at $\omega_{1}$ ($\omega_{2}$) and $\Delta\beta_{nl}\approx\gamma(P_1-P_2)$ for two
pumps in the same band~\cite{ref:Uesaka_Kazovksy}, we use the same design
parameters as~\cite{ref:Agha_OL} in our device fabrication. We produce 18 mm long, 550~nm x 1200~nm (height x width)
Si$_3$N$_4$-on-SiO$_2$ waveguides with an air top cladding, and can adjust the ratio of pump powers
to eliminate the nonlinear phase mismatch. In this experiment, a straight waveguide was employed, however, mature fabrication techniques (both in writing and etching the structure) and the index contrast between Si$_3$N$_4$ and SiO$_2$ allow for integrating the same length in a much smaller ($\approx 1$ mm) footprint via a spiral layout.

\section{Experimental results - Wideband Frequency Downconversion}

\begin{figure}[h!]
\begin{center}
\centerline{\includegraphics[width=\linewidth]{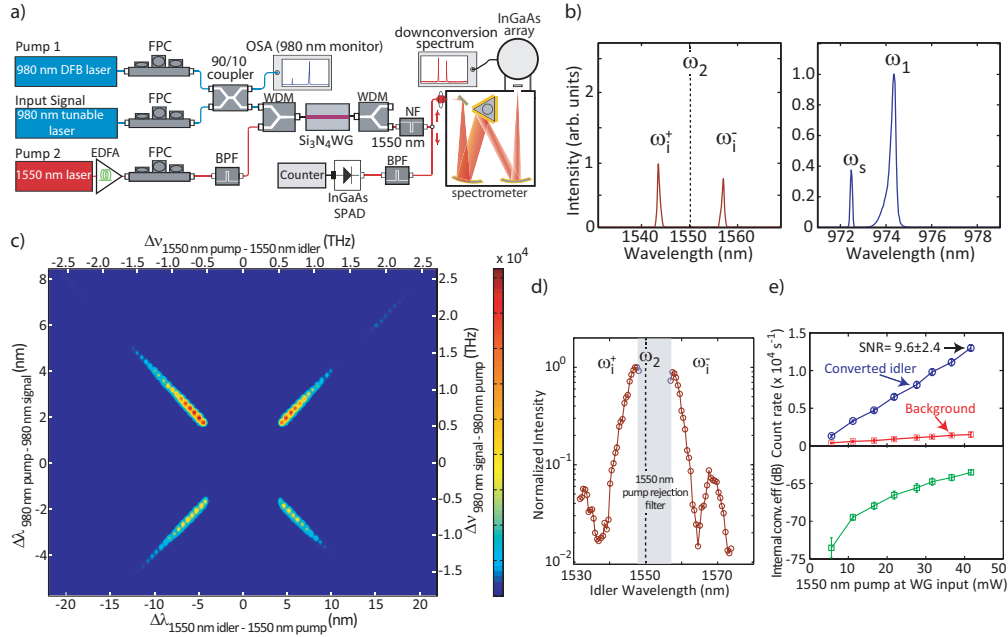}} \caption{a) Experimental setup for frequency downconversion from 980~nm to 1550~nm:
A tunable 980~nm signal laser is combined with a strong 974~nm distributed feedback (DFB) pump laser and then
multiplexed (through a fiber 980~nm/1550~nm wavelength-division multiplexer (WDM)) with a 1550~nm  laser that is
amplified via an erbium-doped fiber amplifier (EDFA). Fiber polarization controllers (FPCs) ensure that the
signal and pumps are co-polarized and launched into the desired waveguide mode. An optical spectrum analyzer (OSA)
monitors the 980~nm wavelengths, while the output of the waveguide is first filtered through a notch filter (NF) to remove the residual 1550~nm pump,
and then either sent to an InGaAs spectrometer or through a bandpass filter (BPF) to an InGaAs single photon
avalanche diode (SPAD). b) Sample spectrum of the generated idlers together with an OSA trace of the 980~nm-band pump+signal.  The position of the 1550~nm pump, which has been notch filtered, is denoted by a dashed line. c) Contour plot of the generated 1550~nm idler spectra (in units of detuning from the 1550~nm pump) for differing levels of 980~nm band pump-input signal detuning (taken with 0.2~nm tuning steps). d) Normalized idler power as a function of its wavelength for fixed pumps and tunable 980~nm input signal. The gray central region denotes the bandwidth of the 1550~nm pump rejection notch filter. e) Conversion efficiency (bottom) and converted idler and background count rate (top) as a function of the 1550~nm pump power inside the waveguide at its input. The error bars (often smaller than the data point size) are due to the variation in the detected SPAD count rates, and are one standard deviation values. The signal-to-background level for the converted idler increases from 5.6$~\pm~$2.5 at the lowest pump power to 9.6$~\pm~$2.4 at the highest pump power.} \label{fig:downconversion_data}
\end{center}
\end{figure}

The experimental setup for frequency downconversion of a 980~nm band signal
to a telecom--band idler is shown in Fig.~\ref{fig:downconversion_data}(a). A
strong (550 mW) pump consisting of a 974~nm distributed feedback laser is
combined with a weak, tunable signal on a 90:10 coupler, and the combined
output is sent to the 980~nm port of a 980~nm/1550~nm wavelength-division
multiplexer (WDM). The signal power at the input of the waveguide is in
the range of a few $\mu W$. The second pump is a 1550~nm laser that is amplified via
an erbium-doped amplifier (EDFA) and sent to the other port of the WDM. The
combined pumps and signal are coupled to the waveguide via a lensed fiber,
and the output of the waveguide is filtered to reject both the residual
(unconverted) signal and the pumps. This output is either sent to  a
spectrometer equipped with a cooled InGaAs photodiode array for spectral analysis, or is
further bandpass-filtered and sent to an InGaAs single photon avalanche diode
(SPAD) for signal-to-noise analysis. Data is synchronously recorded on the
spectrometer while the 980~nm band signal laser is tuned from 965~nm to
985~nm in 0.2~nm steps and monitored on an optical spectrum analyzer (OSA).

Figure~\ref{fig:downconversion_data}(b) shows the generated idlers in the 1550~nm band for a specific set of pump and input signal wavelengths in the 980~nm band (the 1550~nm pump has been removed by a notch filter after the WG chip, as shown in Fig.~\ref{fig:downconversion_data}(a)).  Tuning the signal away from the 974~nm pump causes the idlers to tune in accordance with energy conservation $\omega_i=\omega_2 \pm (\omega_s - \omega_1)$, as shown in the image plot in Fig.~\ref{fig:downconversion_data}(c).  Here, the x-axis is in units of detuning between the generated idlers and the fixed 1550~nm pump, while the y-axis is in units of detuning between the input 980~nm band signal and fixed 974~nm pump.

Figure~\ref{fig:downconversion_data}(d) shows the normalized conversion
efficiency as a function of the idler wavelength (for fixed pump
wavelengths). This plot shows both the phase matching bandwidth of our device
as well as  oscillations which are likely due to its expected sinc$^2$
character (Eq. 1). We next use an InGaAs SPAD to determine both the
conversion efficiency as well as the signal-to-background ratio for the converted idler as a function
of the 1550~nm pump power. The signal-to-background is determined by
measuring the counts in the $\omega_{i}^{+}$ idler band with the input 980~nm
signal on and off, and dividing the two values (after subtraction of the detector
dark count rate of $\approx 150$~s$^{-1}$). As the pump power inside the waveguide and at its input is
increased from 5~mW to 43~mW, the internal conversion efficiency (taking into
account 7.5~dB input and output coupling losses) increases from -74 dB to -64
dB while the signal-to-background increases from $\approx 5$ to $\approx 10$.
While nearly background-free FWM-BS conversion has been demonstrated in
situations for which the detuning between pumps and signal and idler is
large~\cite{ref:McGuinnes_PRL10}, for smaller detuning levels, high
signal-to-background levels are generally harder to achieve. This can be due to
technical reasons, such as incomplete suppression of the pump at the
converted idler wavelength, or more fundamental reasons related to the
waveguide system itself, such as spontaneous Raman scattering in
SiO$_2$~\cite{ref:clark2013} or modulation instability when the strong pumps are
situated in regions of anomalous dispersion.  Assuming these effects contribute to the measured
noise level, the relatively high signal-to-background levels observed suggest that the
Si$_3$N$_4$ system is less susceptible to noise from Raman scattering than systems such as
silica optical fibers. If these signal-to-background levels can
be maintained at higher conversion efficiency levels, they would be adequate
for initial quantum frequency conversion experiments with true single photon sources
(signal-to-background ratios between 2 and 7 were used in
Ref.~\cite{ref:Rakher_NPhot_2010}).

In moving to such experiments, of principal concern is the very low
conversion efficiencies thus far demonstrated.  However, this experiment was
done under continuous-wave and relatively low pump power conditions, with
maximum powers of $P_1=13$ mW and $P_2=43$ mW at the input inside the
waveguide. These results are similar to those in Ref.~\cite{ref:Agha_OL}
achieved under similar power conditions. In that work, by using amplified
pulses, the conversion efficiency improved to $\approx 5~\%$ for peak powers
of a few Watts coupled into the chip. With similar pump amplification (for
example, using nanosecond pulses and an EDFA to generate the telecom-band
high peak-power pump and a tapered amplifier for the 980~nm band pump), and
by reducing the input coupling losses via optimized geometries (for example,
inverse tapers in symmetrically clad coupling regions) - already
demonstrated for Si$_3$N$_4$ waveguides~\cite{ref:Ferdous_Weiner} - we expect
the conversion efficiency to improve to over $25~\%$, as confirmed by split-step
Fourier method simulations~\cite{ref:Agha_OL}. This conversion efficiency level would
render the interface suitable for quantum frequency conversion with single photon emitters~\cite{ref:Rakher_NPhot_2010,ref:McGuinnes_PRL10}.

\begin{figure}[h!]
\begin{center}
\centerline{\includegraphics[width=\linewidth,trim= 0 0 0
0]{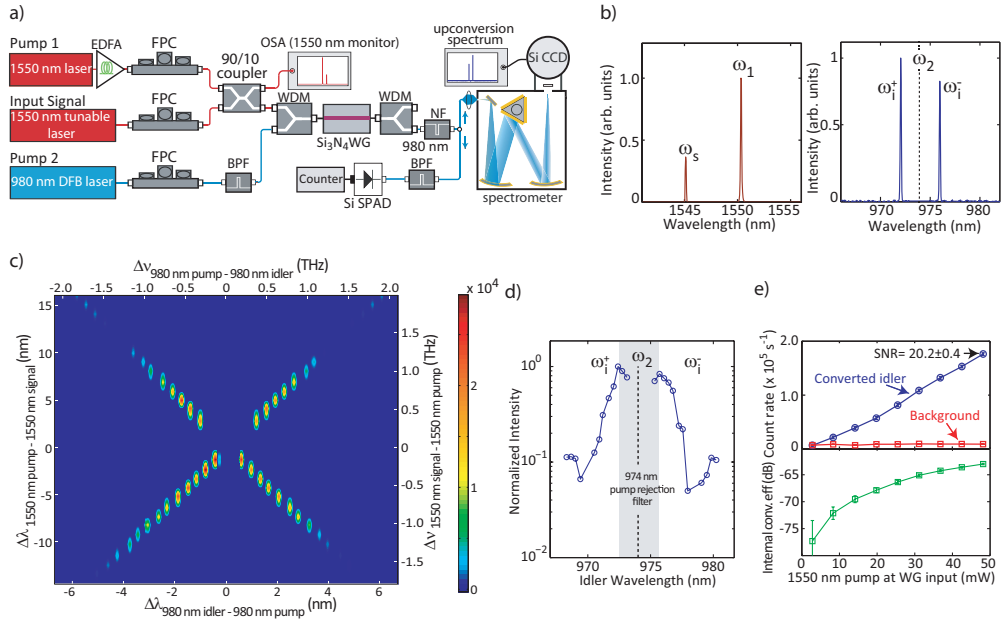}} \caption{a) Experimental setup for frequency
upconversion from 1550~nm to 980~nm:  A tunable 1550~nm signal laser is combined
on a 90:10 coupler with a 1550~nm pump and then multiplexed
(through a fiber 980~nm/1550~nm WDM) with a 974~nm pump laser. The output of the
WDM is sent to the waveguide chip, and the output of the waveguide is notch-filtered to
remove the residual 974~nm pump. The 1550~nm wavelengths are monitored on the
OSA, while the converted 980~nm band idlers are measured on a spectrometer equipped
with a Si CCD, or are bandpass filtered and measured on a Si SPAD. b) Sample
spectrum of the generated idlers together with an OSA trace of the 1550~nm-band
pump+signal.  The position of the 974~nm pump, which has been notch filtered, is denoted by a dashed line. c) Contour plot of the generated 980~nm idler spectra
(in units of detuning from the 974~nm pump) for different levels of 1550~nm band
pump-input signal detuning (taken with 1~nm tunings steps). d) Normalized idler power as a function of its wavelength for fixed pumps and tunable 1550~nm input signal. The gray central region denotes the bandwidth of the 974~nm pump rejection notch filter. e) Conversion
efficiency (bottom) and converted idler and background count rate (top) as a function of the 1550~nm pump power
inside the waveguide and at its input.  The error bars (often smaller than the
data point size) are due to the variation in the detected SPAD count rates,
and are one standard deviation values. The signal-to-background level for the converted idler increases up to 20.2$~\pm~$0.4 at the highest pump power.} \label{fig:upconversion_data}
\end{center}
\end{figure}

\section{Experimental results - Wideband Frequency Upconversion}

The flexibility of the two pump FWM-BS process implies that by switching the
roles of the signal and idler, we can perform frequency upconversion within the same device. To ensure the
capability of single photon counting in the 980~nm band, the 974~nm pump
laser is heavily filtered via volume Bragg gratings to yield better than 140
dB of suppression. The 1550~nm fixed pump laser is combined with a tunable
telecommunications-band weak signal via a 90:10 coupler, and the combined
pumps+signal are multiplexed via a WDM and sent to the waveguide chip
(Fig.~\ref{fig:upconversion_data}(a)). The output of the chip is
notch-filtered (to get rid of the strong 974~nm laser) and sent to either a
visible-wavelength spectrometer or silicon SPAD for signal-to-background
measurements.

Figure~\ref{fig:upconversion_data}(b) shows the generated 980~nm
band idlers for a specific set of pump and input signal wavelengths in the 1550~nm band (the 980~nm band
pump has been removed by a notch filter after the chip, as shown in Fig.~\ref{fig:upconversion_data}(b)).
Combining a series of upconversion spectra as the 1550~nm band input signal is swept yields the contour plot shown in
Fig.~\ref{fig:upconversion_data}(c), which confirms a tuning of the generated idlers that matches the energy conservation condition,
$\omega_i=\omega_s \pm (\omega_1 - \omega_2)$. By integrating the power
within each idler peak, we can deduce the normalized conversion efficiency
and hence the phase-matching bandwidth, which, as expected, is similar in
behavior to that observed in downconversion
(Fig.~\ref{fig:downconversion_data}(d)). As the 1550~nm  pump power is
increased up to 50 mW, the internal conversion efficiency improves from -77
dB to -62 dB, while the signal-to-background ratio for the $\omega_{i}^{+}$
idler reaches a maximum of $\approx 20$ (Fig. 3(e)).  As in the case of downconversion,
the signal-to-background ratio is determined by measuring the counts in the
$\omega_{i}^{+}$ band with the input 980~nm signal turned on and off, and dividing the two values
(after subtraction of the detector dark count rate of $\approx 170$~s$^{-1}$). The higher
signal-to-background we measure here for upconversion can most likely be
attributed to the improved filtering of the pumps both at the input and the
output of the waveguide via volume Bragg gratings that provide better
extinction than fiber-based WDMs.  This also provides some indication that
the process may not be limited by noise contributions due to Raman
scattering. In comparison to SiO$_2$, we note that Raman scattering in
silicon nitride has not been observed to be comparable to four-wave-mixing,
even at high power
conditions~\cite{ref:Levy_Lipson_comb,ref:Ferdous_Weiner,ref:Foster_Levy,
ref:Okawachi}.  Another potential reason for the higher signal-to-background levels
is that the strong 980~nm pump is situated deep within the normal dispersion region (Fig.~\ref{fig:waveguide and schematics}(b)),
so that modulation instability is completely suppressed.  In comparison, the 1550~nm
pump, though nominally also in a region of normal dispersion, is much closer to the dispersion zero.
As a result, deviations in the waveguide geometry with respect to the simulated structure (e.g., due
to fabrication tolerances or a non-uniform waveguide cross-section along its length) may lead to
modulation instability induced noise.

\section{Experimental results - Narrowband Frequency Conversion}

\begin{figure}[t!]
\begin{center}
\centerline{\includegraphics[width=\linewidth,trim= 0 0 0
0]{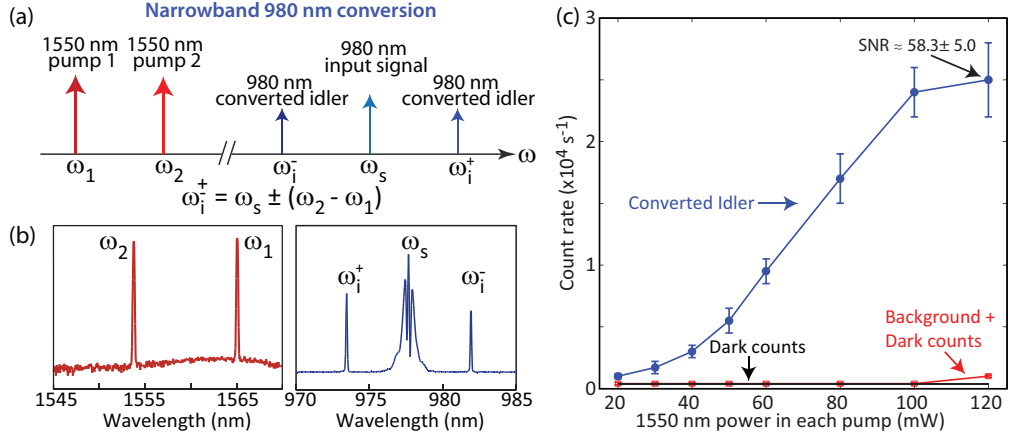}} \caption{a) FWM-BS process for nearly noise-free, narrowband
frequency conversion: two telecommunications-band pumps cause a wavelength-exchange
between a signal and its idler according to the energy conservation rule:
$\omega_i=\omega_s \pm (\omega_2 - \omega_1)$. b) OSA spectrum of the telecommunications-band
pumps and spectrometer record of the generated 980~nm band idlers (residual unconverted signal attenuated via fiber Bragg gratings).
c) Converted idler and the corresponding background vs. waveguide input pump power. For the majority of the
input power range, no excess noise above the SPAD dark count rate is observed.  At the highest
pump power, the signal-to-background ratio is $\approx 50$, and is limited by excesss noise from
the pump EDFA.  The error bars are due to the variation in the detected SPAD count rates,
and are one standard deviation values.} \label{Fig:narrowband_SNR_data}
\end{center}
\end{figure}

As noted previously, FWM-BS in the case of pumps that are far-detuned from
the input signal and generated idlers has likely advantages from a
signal-to-background perspective.  To some extent, it can thus provide a
reference point for the best achievable signal-to-background levels within
the system.  To verify this, we revert to the narrowband conversion
configuration (Fig.~\ref{Fig:narrowband_SNR_data}) we studied recently in
Ref.~\cite{ref:Agha_OL}. Here, two telecommunications-wavelengths pumps
$\omega_1$ and $\omega_2$ convert a 980~nm band input signal $\omega_s$  to an idler $\omega_i$
in the same band, according to the rule $\omega_i=\omega_s \pm (\omega_2 -
\omega_1)$. At optimal conversion conditions, the pumps are separated by
$\approx 10$~nm, while the signal/idler are separated by $\approx 5$~nm
(Fig.~\ref{Fig:narrowband_SNR_data}(b)). We bandpass filter the generated
idler $\omega_{i}^{+}$ and use a Si SPAD to measure the power in this
wavelength band with the input signal turned on and off
(Fig.~\ref{Fig:narrowband_SNR_data}(c)).  When the input is turned off, we
measure no excess background counts above the SPAD dark count rate ($\approx 400$~s$^{-1}$) for the
majority of the 1550~nm pump power range sent into the waveguide (so that
after SPAD dark count subtraction, the process is noise-free). Even at the
very highest pump powers, for which we observe significant excess noise in
our EDFA, we measure signal-to-background levels of $\approx 50$.

\section{Discussion}

We have demonstrated the first steps towards a quantum interface for
ultra-wide band frequency conversion, and shown how it is experimentally
possible to translate photons from 980~nm to 1550~nm and back via FWM-BS in a
CMOS-compatible platform. The goal of this work is to pave the way towards
both high-efficiency devices that render on-chip frequency conversion of
single photon emitters practical, as well as to provide design guidelines for
down(up)conversion of photons from various quantum emitters via FWM-BS. While
our devices targeted the 980~nm band which is the wavelength range of InAs
quantum dot SPSs that we have employed in recent
experiments~\cite{ref:Ates_Srinivasan_PRL, ref:Ates_SR}, the ability to
engineer the dispersion, coupled with the transparency of silicon nitride,
allows for devices that in principle can target most common single photon
emitters.

\begin{figure}[b]
\begin{center}
\centerline{\includegraphics[width=\linewidth]{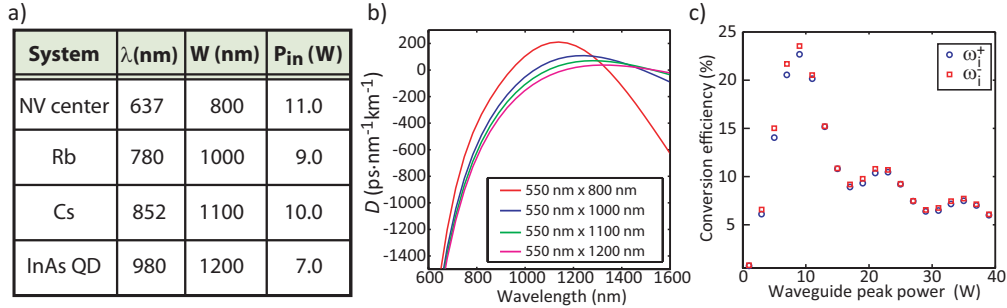}} \caption{a) Design parameters for Si$_3$N$_4$-on-SiO$_2$
waveguides targeting up(down)conversion between the telecommunications band
and various single photon sources (SPSs)  (Nitrogen vacancy center in diamond, Rubidium, Cesium, and
InAs QD in descending order). The table specifies, for each SPS wavelength ($\lambda$),
the waveguide  width (W) and the necessary  input peak power (P$_{in}$) for
maximum conversion.  The waveguide thickness has been fixed at 550~nm in all cases. b) The corresponding
dispersion parameter $D$ for each waveguide geometry. (c) Split-step Fourier simulation for the case of
downconversion between 780~nm and 1550~nm.  Conversion efficiency is plotted as a function of pump power,
where the power in both pumps are assumed to be equal.} \label{fig:Dispersion_table}
\end{center}
\end{figure}

Figure~\ref{fig:Dispersion_table}(a) is a table of common SPS
emission wavelengths, with the required geometry of the waveguide to achieve
phase matching for downconversion towards the 1550~nm telecommunications
band. Along with linear phase-matching, of key importance is for the strong pump
fields to be in regions of normal dispersion, to avoid the onset of modulation instability.
This has been achieved here by simply tuning the waveguide width, while keeping
the thickness fixed at 550~nm.  Figure~\ref{fig:Dispersion_table}(b) shows the
resulting dispersion parameter $D=\frac{-2\pi c}{\lambda^2}\frac{d^2\beta}{d\omega^2}$
for the different optimized geometries, where the zero dispersion point, to a
first approximation, is close to the average of the two  pumps' wavelengths
~\cite{ref:Uesaka_Kazovksy}.  Finally, Fig.~\ref{fig:Dispersion_table}(c) shows an
example of the results from a split-step Fourier simulation~\cite{ref:Agrawal_NFO} in which we
take into account the calculated dispersion to assess the efficiency of the frequency conversion process
without the approximations that go into the analytic coupled-mode theory model (e.g., no pump depletion,
no pump mixing, and no degenerate four-wave-mixing).  The figure plots the conversion efficiency as a function
of pump power (assumed to be equal for the two pumps) in the case of downconversion from 780~nm to 1550~nm.
We see that a conversion efficiency of $\approx25~\%$ can be achieved for approximately 9~W of power in each pump.
Simulations reveal that conversion efficiencies $\gtrsim25~\%$ can be achieved for the other sets of wavelengths considered, and
the corresponding pump powers are listed in the table in Fig.~\ref{fig:Dispersion_table}(a). It should be noted that while 9~W is
a relatively high power when considering continuous-wave sources, it is readily achievable by amplifying nanosecond pulses in saturated erbium-doped or semiconductor tapered 
amplifiers, making our chip-scale interface a practical prospect for conversion of single photon states of light from quantum emitters (particularly systems like semiconductor quantum dots, which have characteristic timescales on the order of 1~ns). Moreover, by implementing the structure in Si$_3$N$_4$, peak power is not limited by phenomena such as two-photon absorption, and hence such powers can be coupled to a Si$_3$N$_4$ waveguide without major problems, as observed in experiment for few Watt level pumps in Ref.~\cite{ref:Agha_OL}.  On the other hand, implementation of the four-wave-mixing Bragg scattering process in crystalline silicon waveguides could be of interest, in part due to its higher nonlinear refractive index than Si$_3$N$_4$~\cite{ref:Foster_Gaeta,ref:Lin_Painter_Agrawal}. Such work would be limited to wavelengths $>1000$~nm (as silicon is opaque at shorter wavelengths), and may require long wavelength ($>2000$~nm) pumps to avoid significant nonlinear absorption~\cite{ref:Liu_Green_NatPhoton,ref:Zlatanovic_Radic_NatPhoton}.  Alternately, materials like amorphous silicon, which combine high optical nonlinearity and a broader optical transparency window than silicon (though less than Si$_3$N$_4$) may ultimately prove to be an attractive candidate for four-wave-mixing Bragg scattering within certain wavelength regions~\cite{ref:Wang_Foster_amorphous_Si}.

In conclusion, we have demonstrated a CMOS-compatible chip-scale
interface for ultra-wide band frequency conversion via four-wave-mixing Bragg
scattering in a silicon nitride waveguide, and characterized both its tuning
bandwidth and noise properties. We performed proof-of-principle experiments
demonstrating both downconversion and upconversion between the 980~nm and
1550~nm bands, rendering our chip a promising interface for connecting
self-assembled InAs quantum dots to the telecommunications band. With
improved input coupling and higher input pump peak powers, we expect our
interface to be ready for frequency conversion of true single photon sources,
which will be the target of future work.

\section{Acknowledgments}

The authors thank Yoshitomo Okawachi for helpful comments on this work.
I.A and S.A. acknowledge support under the Cooperative Research Agreement
between the University of Maryland and NIST-CNST, Award
70NANB10H193. The authors also acknowledge the DARPA MESO program for partial support.

\end{document}